\documentclass[twocolumn,showpacs,preprintnumbers,amsmath,amssymb,prl]{revtex4}

\usepackage{graphicx}% Include figure files
\usepackage{dcolumn}% Align table columns on decimal point
\usepackage{bm}% bold math

\newcommand{\Od}{{\cal O}}

\newcommand{\im}{\mbox{Im}\,}
\newcommand{\re}{\mbox{Re}\,}

\newcommand{\be}{\begin{equation}}
\newcommand{\ee}{\end{equation}}
\newcommand{\ba}{\begin{eqnarray}}
\newcommand{\ea}{\end{eqnarray}}

\newcommand{\da}{\dagger}
\newcommand{\ov}{\overline}

\newcommand{\NP}[1]{Nucl.\ Phys.\ {\bf #1}}
\newcommand{\ZP}[1]{Z.\ Phys.\ {\bf #1}}

\newcommand{\PL}[1]{Phys.\ Lett.\ {\bf #1}}

\newcommand{\PR}[1]{Phys.\ Rev.\ {\bf #1}}
\newcommand{\PRL}[1]{Phys.\ Rev.\ Lett.\ {\bf #1}}

\newcommand{\IJmp}[1]{{Int.\ J.\ Mod.\ Phys.\ }{\bf #1}}
\newcommand{\dnote}[1]{}
\newcommand{\gsim}{\raise.3ex\hbox{$>$\kern-.75em\lower1ex\hbox{$\sim$}}}
\newcommand{\lsim}{\raise.3ex\hbox{$<$\kern-.75em\lower1ex\hbox{$\sim$}}}
\begin{document}

%\preprint{hep ph/02XXXXXX}

\title{Thermal rho and sigma mesons from chiral symmetry and unitarity}

\author{A. Dobado$^1$
%\email{malcon@fis.ucm.es}
}
\author{A. G\'omez Nicola$^1$
%\email{gomez@fis.ucm.es}
}
\author{F. J. Llanes-Estrada$^1$
%\email{fllanes@fis.ucm.es}
}
\author{J. R. Pel\'aez$^{1,2}$
%\email{jrpelaez@fis.ucm.es}
}
\affiliation{$^1$Departamentos de
F\'{\i}sica Te\'orica I y II,  Universidad Complutense. 28040 Madrid,
Spain. \\
$^2$Dip. di Fisica. Universita' degli Studi, Firenze,
  Italy and INFN, Sezione di Firenze, Italy}

\begin{abstract}
We study the temperature evolution of the $\rho$ and $\sigma$ mass
and width, using a  unitary chiral approach. The one-loop $\pi\pi$
scattering amplitude in Chiral Perturbation Theory at $T\neq 0$ is
unitarized via the Inverse Amplitude Method. Our results predict a
clear increase with $T$ of both the $\rho$ and $\sigma$ widths.
The masses decrease slightly for high $T$, while the $\rho\pi\pi$
coupling increases. The $\rho$ behavior seems to be favored by
experimental results. In the $\sigma$ case,   it signals chiral
symmetry restoration.
\end{abstract}


\maketitle

One of the outstanding phenomena related to heavy ion collisions
is the flatness of the dilepton spectrum near the  mass of the
$\rho$ meson, which is so clearly visible in many  processes
involving hadrons and electromagnetic probes. This flatness has
been observed by the  HELIOS and CERES collaborations
\cite{helios95,ceres99} and has been the subject of widespread
discussion. Dileptons and photons provide neat signals of the
early stages of the quark-gluon plasma and its  subsequent
evolution into a hadron gas \cite{alam01}. In fact, the most
credible explanation  of the absence of a prominent hill in the
dilepton spectrum is a change in the mass and width  of the $\rho$
due to its interactions with the hot hadron gas
\cite{haglin,li,soko96,ele01}. Since the baryons, with a large
forward momentum, have almost escaped the central collision
region, this gas is composed mainly of pions. Our aim is to study
the thermal evolution of the $\rho$ mass $M_\rho$ and width
$\Gamma_\rho$, from the first principles of chiral symmetry and
unitarity in $\pi\pi$ scattering.

What happens to the $\rho$ in extreme conditions is a hadronic
physics problem, involving non perturbative physics and hence
difficult to be treated. Prior to this work, a copious number of
models and estimations have appeared. In most of them
$\Gamma_\rho$ increases with temperature, simply as a consequence
of stimulated emission
 in the pion thermal bath  or, equivalently,
  because the effective phase space increases \cite{pis95,weldon93}. This behavior is
  often interpreted as a deconfining effect, or hadron "melting".
  As for the mass,  Vector Meson Dominance (VMD) implies that
  $M_\rho$  changes very little at low
  temperatures \cite{de90,GaKa91,pis95}. As $T$ approaches the critical temperature,
 earlier works claimed that $M_\rho$
increases
  \cite{GaKa91,dom93,pis95} but  the
analysis of experimental dilepton data seems to favor a decreasing
 behavior \cite{li,ele01}. Let us remark that in all these
works, the $\rho$ is introduced as an explicit degree of freedom
and often a dilute pion gas is assumed, so that the thermal
effects appear, to leading order, only through the pion
distribution function  and not through the interaction details.
Other approaches include the NJL model \cite{NJL1}, where $M_\rho$
and $\Gamma_\rho$ slightly decrease (but there is an spurious
quark threshold near $M_\rho$) as well as $q\overline{q}$
wave-function  analysis in the $\pi$ channel yielding a decreasing
width  \cite{Maris}.

In this work we will use a thermal treatment of the effective
degrees of freedom, the pions in the aftermath of the collision at
moderate temperatures. The guiding fundamental principles will be
just chiral symmetry and unitarity. We will build on a previous
work \cite{ouramp} where the $T\neq 0$ $\pi\pi$ scattering
amplitude has been calculated to one loop in Chiral Perturbation
Theory (ChPT). Demanding unitarity, we will construct a
non-perturbative amplitude  reproducing the expected behavior for
thermal resonances. Our amplitude has the correct analytic
structure, without spurious cuts, and resonances  are not
introduced by hand.

The most general framework comprising the QCD chiral symmetry
breaking pattern  is ChPT \cite{we79,gale84}, see \cite{books}
for reviews,
 where observables are
calculated as expansions in $p/(4\pi f_\pi)$, $p$ denoting any
pion energy scale (including the temperature) and $f_\pi\simeq$
92.4 MeV. Despite its success, ChPT is limited to low energies
(usually, less than 500 MeV) and low temperatures and it is not
able to generate resonances. Thus, over the last few years, there
has been
 a growing interest to extend the ChPT applicability range   to
higher energies and to reproduce resonances within a unitary
chiral approach, which we briefly review. At $T=0$, unitarity for
the $S$-matrix ($S^\da S=1$)  implies the following relation for
partial waves
\begin{equation}  \label{unit0}
 \im a_{IJ}(s)=\sigma (s) \vert a_{IJ} (s) \vert^2 \ ,
\end{equation}
for $s>4m_\pi^2$ and below other inelastic thresholds, where
$\displaystyle \sigma (s)=\sqrt{1-4 m_\pi^2/s}$ is the two-pion
phase space and $a_{IJ}$ denotes the projection of the $\pi\pi$
elastic amplitude with isospin $I$ and total angular momentum $J$
in the center of mass frame. Eq.(\ref{unit0}) is only satisfied
{\em perturbatively} within ChPT, i.e, if we write the
perturbative series for any partial wave as $a=a_2+a_4+ \dots$
where $a_{k}$ is $\Od(p^k)$, then one has $\im a_2=0$, $\im
a_4=\sigma a_2^2$ and so on. Hence, deviations from
eq.(\ref{unit0}) are more severe at high energies, and in
particular near the resonance region, where the bounds imposed by
unitarity are saturated. The ChPT series, which essentially
behaves as a polynomial, is unbounded and cannot reproduce
resonances, which show up as poles of the amplitude in the complex
plane.

In fact, from eq.(\ref{unit0}), any  partial wave should satisfy
$a=1/(\re a^{-1}-i \sigma)$ on the real axis below inelastic
thresholds. A unitarization method is just one way of
approximating $\re a^{-1}$, thus introducing some
model dependency, but since we want to ensure chiral
symmetry, and the correct low energy behavior at $T=0$,
we will use the one-loop ChPT result. This is called the
IAM  at $T=0$, which can be recast as $a^{IAM}=a_2^2/(a_2-a_4)$
\cite{IAM}. 
The single channel IAM amplitude
satisfies eq.(\ref{unit0}) exactly and at low energies it follows
the ChPT result up to one loop. In addition, the IAM reproduces
the scattering data for real energies above the two pion threshold
up to 1 GeV, where the elastic approximation breaks down, and it
can be continued into the complex $s$-plane, yielding correct
$\sigma$ and $\rho$ poles in the second Riemann sheet. We point
out that the IAM is nothing but the [1,1] Pad\'e approximant of
the ChPT  series in squared energy, mass, or temperature over $f_\pi^2$. 

 As long as they
contain the $\Od(p^4)$ tree level terms,
other chiral unitary approximations, both for SU(2) or SU(3)
ChPT, either based on the IAM with coupled channels \cite{IAMcoupled}, 
or the IAM with higher orders
\cite{JuanIAM}, or inspired in Lippmann-Schwinger
or Bethe-Salpeter equations \cite{LS},
or mixed formalisms \cite{IAM2}
yield equivalent
results  for the $\rho$ and $\sigma$ channels. In particular, 
they reproduce the experimental phase shifts with compatible sets
of chiral parameters, and they generate poles 
associated to the $\sigma$ and $\rho$ resonances whose
position in the second Riemann sheet agrees for all the
above mentioned methods, which therefore describe resonances with the 
same masses and widths.
These unitarized approaches also allow to
 study finite baryon density effects on
the change of the sigma properties in the nuclear medium \cite{EulogioVacas}
that suggested  a   decrease on both the sigma mass and width as the
nuclear density increases. As a consequence of these effects,
it is expected \cite{Eulogio} a shift of strength
of the two pion invariant mass distribution in 
$\gamma N\rightarrow N\pi^0\pi^0$, which has been recently confirmed
experimentally \cite{Mainz}. In particular, using a chiral unitary
approach, this shift is interpreted as an in medium modification of
the $\sigma$ pole towards lower masses and widths
Nevertheless, since in this work we are interested in the $\rho$ and $\sigma$
mesons it is enough to work with the single channel IAM to $O(p^4)$
that we have just described.

Back to $T\neq 0$, the thermal amplitude can be defined by
considering $T=0$ initial and final asymptotic states and
calculating  the $T\neq 0$  four-pion  Green's function
\cite{ouramp}. To one loop in ChPT, and in the $\pi\pi$ c.o.m.
frame (at rest  with the thermal bath) it satisfies the {\em
perturbative} unitarity relation \cite{ouramp}:
\begin{equation} \label{unittf}
\im a_4 (s;T)=\sigma_T (s) \left[a_2(s)\right]^2
\end{equation}
where
\begin{equation} \label{thps}
 \sigma_T(s)=\sigma (s)\left[1+2
n_B\left(\sqrt{s}/2\right)\right]
\end{equation}
is the thermal phase space  and $n_B(x)=(\exp(x/T)-1)^{-1}$ is the
Bose-Einstein distribution function. Recall that the lowest order
$a_2$ is $T$-independent.

Therefore, the natural unitarized version of the
thermal amplitude in ChPT should be:
\begin{equation} \label{theriam}
a^{IAM}(s;T)=\frac{a_2^2(s)}{a_2(s)-a_4(s;T)}
\end{equation}  which satisfies the exact elastic unitarity
condition
\begin{equation} \label{iamunittf}
\im a^{IAM} (s;T)=\sigma_T (s) \left\vert
a^{IAM}(s;T)\right\vert^2
\end{equation} and  reproduces
the low energy results of (thermal) ChPT  in \cite{ouramp}.
Besides, as we will see below,  it has the proper analytical
behavior and, for the appropriate values of the chiral
parameters, it is able to reproduce resonances like the $\rho$ as
poles in the second Riemann sheet.

Some remarks are in order here: We are assuming that the  exact
thermal version of eq.(\ref{unit0}) holds, feature reproduced by
the IAM in eq.(\ref{iamunittf}).
 This assumption will prove to be  reasonable
in view of the results shown below. Nevertheless, it is important
to remark that such assumption implies in particular that only
two-pion states are available in the thermal bath. This is
equivalent to a dilute gas approximation. In other words, the
$n_B$ term in eq.(\ref{thps}) must remain small compared to one so
that we can neglect higher orders in density like $\Od(n_B^2)$
which would spoil the simple algebraic unitarity relation given by
eq.(\ref{unittf}) \cite{ouramp}. This implies, for instance that
$\rho-\pi$ scattering, which in our approach is regarded as a
three-pion effect, would be suppressed by the low density. Note
that, alternatively, we can view eq.(\ref{theriam}) also as the
[1,1] Pad\'e approximant of ChPT when counting the powers
of momenta, masses or temperature over 
$1/f_\pi^2$, since
$T$ is $\Od(p)$ in the chiral expansion. Again, this counting
would be spoiled for large $n_B(\sqrt{s})$, which typically
weights the thermal corrections.  Let us finally remark that the
IAM has been extended to deal with other intermediate states,
describing successfully the $T=0$ data in all meson-meson channels
up to 1.2 GeV, within a coupled channel formalism
\cite{IAMcoupled}. In such case, the amplitudes satisfy a matrix
version of the unitarity relation in  eq.(\ref{unit0}). This would
be the natural extension of our approach in order to deal with
other intermediate coupled states, like $K$ or $\eta$ which could
be relevant at high temperatures \cite{GeLe}. However, it would
require a thermal generalization of the matrix unitarity relation
and the one-loop calculation of the additional coupled amplitudes,
which lie beyond the scope of this work.

Before proceeding to the detailed calculation of the IAM thermal
amplitude in eq.(\ref{theriam}), we will provide a   simple
argument as to why our method can actually give rise to  the
expected thermal behavior for the $\rho$ width. As it is well
known, in most cases (like the $\rho$)  a resonant behavior can be
reproduced on the real axis by means of  a Breit-Wigner
parameterization of the partial waves:
\begin{equation} \label{brewig}
a^{BW}(s;T)=\frac{R_T(s)}{s-M_T^2+i\Gamma_TM_T}
\end{equation}
where $M_T$ and
 $\Gamma_T$ are the thermal mass and width of the resonance
 \cite{weldon93} and $R_T(s)$ is a  smooth  real function
 near $s=M_T^2$, which can be related to the $\rho\pi\pi$ vertex (see below).
    The  parameterization in eq.(\ref{brewig}) applies only for
$s\simeq M_T^2$ and for narrow resonances ($\Gamma_T\ll M_T$).
  Comparing  eq.(\ref{theriam}) with eq.(\ref{brewig}) at $s=M_T^2$ one
readily gets $\re a_4 (M_T^2)=a_2(M_T^2)$ (the resonance mass
condition) and, using eq.(\ref{iamunittf}),
$\Gamma_T M_T=-R_T (M_T^2) \sigma_T (M_T^2)$.
Therefore,  assuming that the thermal
corrections to $R_T$  and to $M_T$  are much smaller than those to
$\Gamma_T$, i.e, $R_T \simeq R_0$ and $M_T\simeq M_0$ we would get
\begin{equation} \label{gammaphase}
\Gamma_T\simeq \Gamma_0 \left[1+2 n_B\left(M_0/2\right)\right]
\end{equation}
Hence, in this limit the thermal IAM yields an increasing
resonance width driven only by the available thermal phase space
eq.(\ref{thps})  for a $\rho$ at rest \cite{pis95,weldon93}. The
above result takes into account the stimulated emission
$\rho\rightarrow\pi\pi$ and absorption $\pi\pi\rightarrow\rho$
from the thermal bath \cite{ouramp} and gives the dominant effect
at very low temperatures, as our full analysis below confirms.
This approximation indicates that the unitarity requirements on
the amplitude capture the qualitative thermal resonance behavior.
Note that, from the resonance mass condition, taking $M_T\simeq M_0$
is equivalent to ignoring the $T$-dependence in $\re a_4 (s;T)$.

Therefore,  by using the full thermal  amplitude $a_4 (s;T)$ in
\cite{ouramp}, we will calculate below both the $M_T$ corrections
and the deviations from eq.(\ref{gammaphase}). Moreover we will
find the analytic continuation of the amplitude to the complex
plane, so that we can describe the resonances as poles of the
thermal amplitude. This is particularly important for the
$\sigma$, whose description  in terms of eq.(\ref{brewig}) is not so
appropriate due to its large width.

Let us note that the breaking of Lorentz invariance of the thermal
formalism allows for a definition of a ``transversal'' and a
``longitudinal'' mass, however, in our case, since
 we are working in the c.o.m. frame, where the $\rho$
is at rest with the thermal bath, both mass definitions coincide
\cite{GaKa91}.

In the c.o.m. frame,  the thermal one-loop amplitude
can be written in terms of the loop functions
\cite{ouramp} \footnote{In the notation of \cite{ouramp},
$\Delta J_0^s (s)$ and $\Delta J_{0,2}^{tu} (t)$ correspond to
$\Delta J_0 (\sqrt{s},\vec{0})$ and
$\Delta J_{0,2} (0,\sqrt{-t})$ respectively.}:
\begin{eqnarray} \label{j0s}
\Delta J_0^s (s;T) \! \!  &=& \! \! \! \! -\frac{1}{\pi^2}\int_{m_\pi}^\infty \! dE
\frac{\sqrt{E^2-m_\pi^2} n_B(E)}{s-4E^2}  \\
\Delta J_0^{tu}(t;T) \! \! &=& \! \! \! \! \frac{1}{4\pi^2 \sqrt{-t}}
\int_0^\infty \! \! \! \! dq \frac{q n_B(E_q)}{E_q}\log\left\vert
\frac{2q+\sqrt{-t}}{\sqrt{-t}-2q}\right\vert \nonumber\\
\Delta J_2^{tu}(t;T) \! \! &=& \! \! \! \! \frac{1}{4\pi^2\sqrt{-t}}
\int_0^\infty \! \! \! \! dq q E_q n_B(E_q)\log\left\vert
\frac{2q+\sqrt{-t}}{\sqrt{-t}-2q}\right\vert \nonumber
\end{eqnarray}
for real $s> 4m_\pi^2$ and real $t<0$,
where $\Delta F(T)\equiv F(T)-F(0)$, $E_q^2=q^2+m_\pi^2$,
   $J_0^s(s;0)$ is given in \cite{gale84} (after the standard
      $\overline{MS}-1$ renormalization)
and $J_{0,2}^{tu}(t;0)$ can be written in terms of $J_0^s(t;0)$.
Note that  on the real axis the only imaginary part comes from
$\im J_0^s (s+i\epsilon;T)=\sigma_T(s)/16\pi^2$ for $s>4m_\pi^2$,
thus ensuring eq.(\ref{unittf}) \footnote{Remember that
$t(s,x)=(x-1)(s-4m_\pi^2)/2$,  $u(s,x)=t(s,-x)$ with $x$  the
cosine of the scattering angle.}.

\begin{figure}
\includegraphics[scale=.56]{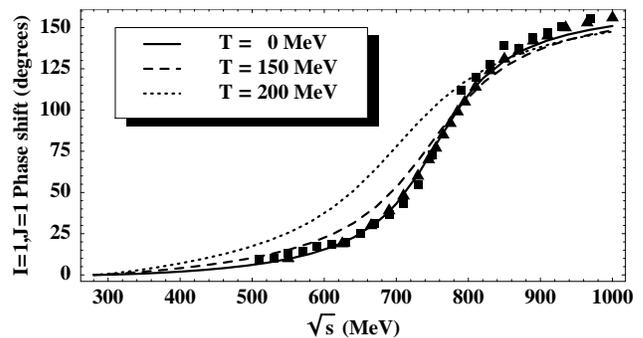}% Here is how to import EPS art
\vspace{-.3cm}
\caption{\rm \label{fig:phase}  $I=J=1$ phase shift for different
temperatures. For the data see \cite{IAM} and references therein.}
\end{figure}

\begin{figure}
\includegraphics[scale=.44]{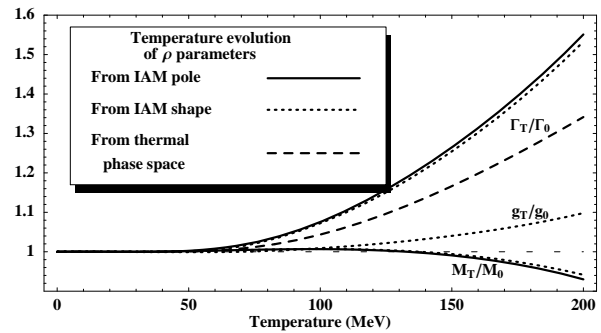}% Here is how to import EPS art
\vspace{-.3cm}
\caption{\rm \label{fig:rhomw} Temperature evolution of the $\rho$
mass, width and  $\rho\pi\pi$  coupling. The dashed line
corresponds to eq.(\ref{gammaphase}), the dotted line to the  real
axis IAM,
 eq.(\ref{theriam}), and the solid line is the IAM pole position.
The $M_0,\Gamma_0,g_0$ values are given in the text.}
\end{figure}

We have calculated the IAM amplitude, eq.(\ref{theriam}), from the
one-loop thermal $a_4(s;T)$. The phase shifts for different
temperatures are shown for the $\rho$ channel $I=J=1$ in
Fig.\ref{fig:phase}. Note the excellent agreement with scattering
data at $T=0$, where we have fitted the $SU(2)$ low-energy
constants  in the $a_{00}$, $a_{11}$, $a_{20}$ channels, yielding
the following values for the standard dimensionless and scale
independent SU(2)
chiral parameters defined in \cite{gale84,books}
$\overline{l}_1=-0.3$, $\ov{l}_2=5.6$, $\ov{l}_3=3.4$ and
$\ov{l}_4=4.3$, compatible with the recent determination in \cite{Amoros}.  
The $\rho$ mass and width can now be estimated
using $\delta_{11}(M_\rho)=90^o$ and $\Gamma_\rho(s)\simeq
M_\rho(1-s/M^2_\rho)\hbox{tan}\delta_{11}$ near $s=M_\rho^2$
\cite{IAM}. Thus we obtain $M_0=$ 770 MeV and $\Gamma_0=$ 159 MeV.
All finite $T$ results are now predictions. As $T$ increases,
$\Gamma_T$ grows, as shown in Fig.\ref{fig:rhomw}. The curves are
shown only below the validity limit of our approach  which naively
is set by $2 n_B(M_\rho/2)< 1$ yielding $T< 300$ MeV. We remark
that the validity of one-loop $SU(2)$ ChPT has been estimated to
reach about $T\simeq$ 150 MeV
 \cite{GeLe},  but  with the unitarization methods we are able to reach
higher temperatures, as long as the density factors remain small
(see our previous discussion).
 To lie on the conservative side, we are only showing results up to
200 MeV, where $2 n_B(M_\rho/2)\simeq$ 0.3. Let us still note that
deviations from the naive phase space correction in
eq.(\ref{gammaphase}) are clearly sizable already at
 $T\simeq$ 100 MeV, precisely when  thermal effects start being
 significant, the full calculation giving a higher value for the width than
eq.(\ref{gammaphase}). The mass changes  little up to $T\simeq$
200 MeV, consistently with previous analysis
\cite{de90,pis95,dom93,ele01,NJL1,Maris}. It grows slightly up to
$T\simeq$ 100 MeV ($M_{100}\simeq $ 775.5 MeV) and then decreases
for higher $T$. In addition, in the narrow resonance approximation
(which becomes less reliable as $T$ increases) we have $R_T= g_T^2
\left(4m_\pi^2-M_T^2\right)/48\pi$, $g_T$ being the
  effective   coupling in the VMD $\rho\pi\pi$ vertex \cite{soko96,pis95}
  with a thermal $\rho$ ($g_0\simeq$ 6.2). Therefore,
  from the IAM $\Gamma_T,M_T$ we find the behavior of
   $g_T$ plotted in Fig.\ref{fig:rhomw}. At low $T$,
$g_T\lsim g_0$ ($g_{50}/g_0\simeq$ 0.9991) in agreement with the
chiral low-$T$  analysis in \cite{soko96} and it grows for higher
$T$. The corrections are more important at finite density
\cite{broflohi01}.

 Although the direct experimental measurement is the dilepton
spectrum, $\pi\pi$ scattering is still a very interesting process
since it is strongly constrained by unitarity. In particular
this provides relevant information about the $\rho$ pole position, which
has to be a common feature for all other processes where the
$\rho$ resonance appears.
That is why we now turn to study the analytic
continuation of the amplitude to the complex plane.
The analytic continuation of the $T=0$  $J_0^s$ is straightforward
 \cite{IAM,gale84}. However, due to the loss of Lorentz covariance
in the thermal bath, we need the analytic continuations of
eqs.(\ref{j0s}) which are somewhat more subtle.
Since $\Delta J_0^s (s;T)$
is already written as an analytic function for $\im
s\neq 0$,  the same expression is straightforwardly continued to
the complex plane. However, for the others we find:
\begin{widetext}
\ba
   \Delta^{\pm} J_0^{tu}(t;T)&=&\frac{1}{4\pi^2 \sqrt{-t}}\left\{
   \int_0^\infty dq \frac{q n_B(E_q)}{E_q}\log\left[
   \frac{2q+\sqrt{-t}}{\sqrt{-t}-2q}\right]
   \pm i\pi T\log\left(1-e^{-R(t)/T}\right)\right\}
   \label{j0tcomp}
\ea
\ba
\Delta^{\pm} J_2^{tu}(t;T)&=&\frac{1}{4\pi^2 \sqrt{-t}}\left\{
   \int_0^\infty dq q E_q n_B(E_q) \log\left[
   \frac{2q+\sqrt{-t}}{\sqrt{-t}-2q}\right]
   \pm i\pi T\left[R^2(t)\log\left(1-e^{-R(t)/T}\right)
\right.\right.\nonumber\\
   &-&\left.\left.2T R(t)
   \mbox{Li}_2\left(e^{-R(t)/T}\right)-2T^2 \mbox{Li}_3\left(e^{-R(t)/T}\right)\right]\right\}
   \label{j2tcomp}
   \ea
   \end{widetext}
where $R(t)=\sqrt{m_\pi^2-t/4}$, $\Delta^{+(-)}$ denote the
analytic continuation for $\im t>0 (<0)$ and Li$_n (z)$ is the
polylogarithmic function, analytic except for a branch cut for
real $z>1$ \cite{erd81}.
It is not difficult to check that $\Delta J_{0,2}^{tu} (t;T)$
coincide with eq.(\ref{j0s}) at $t\pm i\epsilon$ with real $t<0$ and
they have a branch cut only for real $t>4m_\pi^2$.
Thus, as it happened for $T=0$, both $a_4$ and $a^{IAM}$ have a right
(unitarity) cut for real $s>4m_\pi^2$ and a left cut for $s<0$
 coming, respectively, from $\Delta J_0^s$ and
$\Delta J_{0,2}^{tu}$
\footnote{For $s<0$ there is always a finite region in
$x\in[-1,1]$ such that $t,u(s,x)>4m_\pi^2$.}.
 Finally, using eq.(\ref{iamunittf}), the analytic
continuation of the amplitude $a^{II}$ into the second Riemann
sheet across the right cut is given by $a^{II}(s;T)=a^{IAM}
(s;T)/[1-2i \sigma_T (s) a^{IAM} (s;T)]$.

\begin{figure}
\includegraphics[scale=.56]{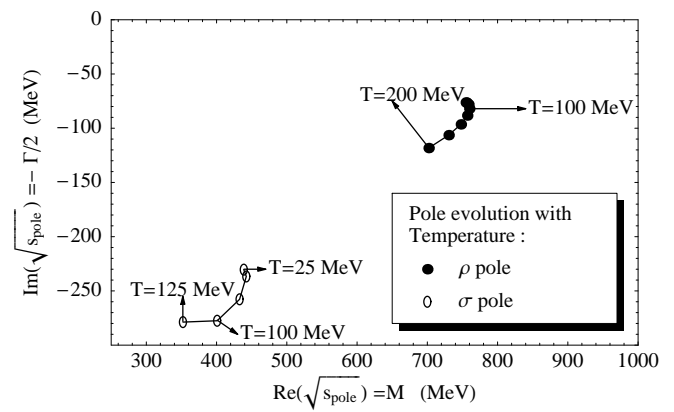}% Here is how to import EPS art
\caption{\rm \label{fig:poles} Position of the $\rho$ and $\sigma$
poles in the complex plane, with increasing temperature.}
\end{figure}

For  $I=J=1$, we find the pole corresponding to the $\rho$
resonance. Its position on the complex plane as a function of $T$
is shown in Fig.\ref{fig:poles}. Let us recall that the
definition of the pole position  in terms of the resonance mass
and width is $s_{pole}=\left(M-i\,\Gamma/2\right)^2$, which
coincides with the pole of eq.(\ref{brewig}) for a narrow
resonance. In particular, at $T=0$, we have $M_0=$ 755 MeV and
$\Gamma_0=$ 152 MeV.  The results  are also plotted in
Fig.\ref{fig:rhomw}, where we see that the evolution of the pole
mass and width agrees with our previous real axis calculation. For
$I=J=0$,  the observed pole corresponds to the $\sigma$ and is
 plotted in Fig.\ref{fig:poles} too. The width also increases,
essentially by the increase of phase space and $M_\sigma (T)$
decreases with $T$, as expected from chiral symmetry restoration
\cite{boka96}. Once again, the applicability of our approach is
limited by $2 n_B(M_\sigma/2)\simeq 1$, i.e, $T<$ 180 MeV.

The main conclusions of this work are the following. We have
shown, using only  chiral symmetry and unitarity, that the thermal
width of the $\rho$ and $\sigma$ mesons at rest with the thermal
bath grow with  temperature, while their thermal masses decrease
slightly. They can be read off  from the real and imaginary parts
of the pole position of the thermal $\pi\pi$ elastic scattering
amplitude in the corresponding channels. For that purpose, we have
  unitarized and calculated the analytic continuation to the complex plane of the
amplitude on the real axis above threshold analyzed in
\cite{ouramp}. For the case of the $\rho$ we have also estimated
its thermal mass, width and  effective $\pi\pi$ coupling from the
 unitarized amplitude in the real axis. At low temperatures, the thermal widths increase
slightly according to the thermal phase space, while the masses
and the effective $\rho\pi\pi$ vertex remain almost constant.
 For higher $T$, our analysis gives sizable decreasing mass corrections, an increasing
  effective vertex,  as
well as significant deviations from the phase space contribution,
yielding higher thermal widths. The $\sigma$ mass shows a
decreasing behavior compatible with chiral symmetry restoration.
Our results agree with recent theoretical and experimental
analysis, up to temperatures of 250 MeV and they shed light on the
dilepton spectrum problem in Relativistic Heavy Ion Collisions.

\acknowledgments Work supported by the Spanish CICYT projects,
FPA2000-0956, PB98-0782 and BFM2000-1326. J.R.P. acknowledges
support from the CICYT-INFN collaboration grant 003P 640.15 and E.
Oset for useful comments.

\end{document}